\input harvmac.tex
\input epsf


\def\Title#1#2{\rightline{#1}\ifx\answ\bigans\nopagenumbers\pageno0\vskip1in
\else\pageno1\vskip.8in\fi \centerline{\titlefont #2}\vskip .5in}
 
scaled\magstep3 
 
scaled\magstep3 
 
scaled\magstep3

\input epsf
\ifx\epsfbox\UnDeFiNeD\message{(NO epsf.tex, FIGURES WILL BE
IGNORED)}
\def\figin#1{\vskip2in}
\else\message{(FIGURES WILL BE INCLUDED)}\def\figin#1{#1}\fi
\def\ifig#1#2#3{\xdef#1{Fig.\the\figno}
\goodbreak\topinsert\figin{\centerline{#3}}%
\smallskip\centerline{\vbox{\baselineskip12pt
\advance\hsize by -1truein\noindent{\bf Fig.~\the\figno:} #2}}
\bigskip\endinsert\global\advance\figno by1}

\baselineskip14pt

\def\alphaprime{\alpha'}
%
%
\def\[{\left [}
\def\]{\right ]}
\def\({\left (}
\def\){\right )}

\def\CO{{\cal O}}

\def\len{{\cal L}}

\def\half{{1 \over 2}}

\def\12{{1 \over 2}}

\def\AS5{{\rm Schw-AdS_5}}


\lref\KloschQV{ T.~Klosch and T.~Strobl,
Universal Coverings,'' Class.\ Quant.\ Grav.\  {\bf 13}, 2395
(1996) [arXiv:gr-qc/9511081].
}

\lref\LeviCX{ T.~S.~Levi and S.~F.~Ross, ``Holography beyond the
horizon and cosmic censorship,'' arXiv:hep-th/0304150.
}

\lref\Baker{ G.A. Baker and P.R. Graves-Morris, { \it Pad{\'e}
Approximants }, second edition, Cambridge Univ. Press,  (1995).}

\lref\IchinoseRG{ I.~Ichinose and Y.~Satoh, ``Entropies of scalar
fields on three-dimensional black holes,'' Nucl.\ Phys.\ B {\bf
447}, 340 (1995) [arXiv:hep-th/9412144].
}

\lref\ShenkerXQ{ S.~H.~Shenker, ``Another Length Scale in String
Theory?,'' arXiv:hep-th/9509132.
}

\lref\KabatCU{ D.~Kabat and P.~Pouliot, ``A Comment on Zero-brane
Quantum Mechanics,'' Phys.\ Rev.\ Lett.\  {\bf 77}, 1004 (1996)
[arXiv:hep-th/9603127].
}

\lref\DouglasYP{ M.~R.~Douglas, D.~Kabat, P.~Pouliot and
S.~H.~Shenker, ``D-branes and short distances in string theory,''
Nucl.\ Phys.\ B {\bf 485}, 85 (1997) [arXiv:hep-th/9608024].
}

\lref\GaiottoRM{ D.~Gaiotto, N.~Itzhaki and L.~Rastelli, ``Closed
strings as imaginary D-branes,'' arXiv:hep-th/0304192.
}

\lref\LambertZR{ N.~Lambert, H.~Liu and J.~Maldacena, `Closed
strings from decaying D-branes,'' arXiv:hep-th/0303139.
}

\lref\MaldacenaBW{ J.~M.~Maldacena and A.~Strominger, ``AdS(3)
black holes and a stringy exclusion principle,'' JHEP {\bf 9812},
005 (1998) [arXiv:hep-th/9804085].
}

\lref\GubserNZ{ S.~S.~Gubser, I.~R.~Klebanov and A.~A.~Tseytlin,
``Coupling constant dependence in the thermodynamics of N = 4
supersymmetric Yang-Mills theory,'' Nucl.\ Phys.\ B {\bf 534}, 202
(1998) [arXiv:hep-th/9805156].
}

\lref\GrisaruZN{ M.~T.~Grisaru, R.~C.~Myers and O.~Tafjord, ``SUSY
and Goliath,'' JHEP {\bf 0008}, 040 (2000) [arXiv:hep-th/0008015].
}

\lref\McGreevyCW{ J.~McGreevy, L.~Susskind and N.~Toumbas,
``Invasion of the giant gravitons from anti-de Sitter space,''
JHEP {\bf 0006}, 008 (2000) [arXiv:hep-th/0003075].
}

\lref\NunezEQ{ A.~Nunez and A.~O.~Starinets, ``AdS/CFT
correspondence, quasinormal modes, and thermal correlators in  N =
4 SYM,'' arXiv:hep-th/0302026.
}

\lref\realtime{For a review see M. Le Bellac, {\it Thermal Field
Theory}, Cambridge University Press (1996).}
\lref\SusskindKW{ L.~Susskind, ``The anthropic landscape of string
theory,'' arXiv:hep-th/0302219.
}

\lref\BerkoozJE{ M.~Berkooz, B.~Craps, D.~Kutasov and G.~Rajesh,
``Comments on cosmological singularities in string theory,'' JHEP
{\bf 0303}, 031 (2003) [arXiv:hep-th/0212215].
}

\lref\GiveonGE{ A.~Giveon, E.~Rabinovici and A.~Sever, ``Beyond
the singularity of the 2-D charged black hole,''
arXiv:hep-th/0305140.
}
\lref\GiveonGB{ A.~Giveon, E.~Rabinovici and A.~Sever, ``Strings
in singular time-dependent backgrounds,'' arXiv:hep-th/0305137.
}

\lref\LoukoTP{ J.~Louko, D.~Marolf and S.~F.~Ross, ``On geodesic
propagators and black hole holography,'' Phys.\ Rev.\ D {\bf 62},
044041 (2000) [arXiv:hep-th/0002111].
}

\lref\etinf{For reviews see A. Linde,
                 {\it Particle Physics and Inflationary Cosmology}
                 Harwood Academic, (1990), and  A.~H.~Guth,
``Inflation and eternal inflation,'' Phys.\ Rept.\  {\bf 333}, 555
(2000) [arXiv:astro-ph/0002156]}

\lref\BerryMI{ M. V. Berry,
 ``Uniform asymptotic smoothing of
Stokes's discontinuities," Proc.\ Roy.\ Soc.\ Lond., {\bf A422}, 7
(1989)~.}

\lref\BerryMII{ M. V. Berry and C. J. Howls,
 ``Hyperasymptotics for integrals
  with saddles," Proc.\ Roy.\ Soc.\ Lond., {\bf A434}, 657 (1991)~.}

\lref\GrossGK{ D.~J.~Gross and H.~Ooguri, ``Aspects of large N
gauge theory dynamics as seen by string theory,'' Phys.\ Rev.\ D
{\bf 58}, 106002 (1998) [arXiv:hep-th/9805129].
}

\lref\StephensAN{ C.~R.~Stephens, G.~'t Hooft and B.~F.~Whiting,
``Black hole evaporation without information loss,'' Class.\
Quant.\ Grav.\  {\bf 11}, 621 (1994) [arXiv:gr-qc/9310006].
}

\lref\SusskindIF{ L.~Susskind, L.~Thorlacius and J.~Uglum, ``The
Stretched horizon and black hole complementarity,'' Phys.\ Rev.\ D
{\bf 48}, 3743 (1993) [arXiv:hep-th/9306069].
}

%
\lref\KrausIV{ P.~Kraus, H.~Ooguri and S.~Shenker, ``Inside the
horizon with AdS/CFT,'' arXiv:hep-th/0212277.
}

\lref\Inami{ T.~Inami and H.~Ooguri, ``One Loop Effective
Potential In Anti-De Sitter Space,'' Prog.\ Theor.\ Phys.\  {\bf
73}, 1051 (1985).
}

\lref\Burgess{ C.~P.~Burgess and C.~A.~Lutken, ``Propagators And
Effective Potentials In Anti-De Sitter Space,'' Phys.\ Lett.\ B
{\bf 153}, 137 (1985).
}

\lref\MaldacenaHW{ J.~M.~Maldacena and H.~Ooguri, ``Strings in
AdS(3) and SL(2,R) WZW model. I,'' J.\ Math.\ Phys.\  {\bf 42},
2929 (2001) [arXiv:hep-th/0001053]; Phys.\ Rev.\ D {\bf 65},
106006 (2002) [arXiv:hep-th/0111180]; J.~M.~Maldacena, H.~Ooguri
and J.~Son, ``Strings in AdS(3) and the SL(2,R) WZW model. II:
Euclidean black
J.\ Math.\ Phys.\  {\bf 42}, 2961 (2001) [arXiv:hep-th/0005183];
J.~M.~Maldacena and H.~Ooguri, ``Strings in AdS(3) and the SL(2,R)
WZW model. III: Correlation functions,'' Phys.\ Rev.\ D {\bf 65},
106006 (2002) [arXiv:hep-th/0111180].
}

\lref\Ben{M. Berkooz, B. Craps, D. Kutasov, and G. Rajesh,
arXiv:hep-th/0212215.}

\lref\Tolley{ A.~J.~Tolley and N.~Turok, ``Quantum fields in a big
crunch / big bang spacetime,'' Phys.\ Rev.\ D {\bf 66}, 106005
(2002) [arXiv:hep-th/0204091].
}

\lref\Herzog{ C.~P.~Herzog and D.~T.~Son, ``Schwinger-Keldysh
Propagators from AdS/CFT Correspondence,'' arXiv:hep-th/0212072.
}


\lref\KutasovXU{ D.~Kutasov and N.~Seiberg, ``More comments on
string theory on AdS(3),'' JHEP {\bf 9904}, 008 (1999)
[arXiv:hep-th/9903219].
}

\lref\deBoerPP{ J.~de Boer, H.~Ooguri, H.~Robins and
J.~Tannenhauser, ``String theory on AdS(3),'' JHEP {\bf 9812}, 026
(1998) [arXiv:hep-th/9812046].
}


\lref\CruzIR{ N.~Cruz, C.~Martinez and L.~Pena, ``Geodesic
Structure Of The (2+1) Black Hole,''
[arXiv:gr-qc/9401025].
}

\lref\MartinecXQ{ E.~J.~Martinec and W.~McElgin, ``Exciting AdS
orbifolds,'' arXiv:hep-th/0206175.
}

\lref\GiveonNS{ A.~Giveon, D.~Kutasov and N.~Seiberg, ``Comments
on string theory on AdS(3),'' Adv.\ Theor.\ Math.\ Phys.\  {\bf
2}, 733 (1998) [arXiv:hep-th/9806194].
}

\lref\SonQM{ J.~Son, ``String theory on AdS(3)/Z(N),''
arXiv:hep-th/0107131.
}

\lref\HananyEV{ A.~Hanany, N.~Prezas and J.~Troost, ``The
partition function of the two-dimensional black hole conformal
JHEP {\bf 0204}, 014 (2002) [arXiv:hep-th/0202129].
}

\lref\TeschnerFT{ J.~Teschner, Nucl.\ Phys.\ B {\bf 546}, 390
(1999) [arXiv:hep-th/9712256];
%
Nucl.\ Phys.\ B {\bf 571}, 555 (2000) [arXiv:hep-th/9906215].
}

\lref\DanielssonZT{ U.~H.~Danielsson, E.~Keski-Vakkuri and
M.~Kruczenski, ``Spherically collapsing matter in AdS, holography,
and shellons,'' Nucl.\ Phys.\ B {\bf 563}, 279 (1999)
[arXiv:hep-th/9905227].
}
\lref\EvansFR{ T.~S.~Evans, A.~Gomez Nicola, R.~J.~Rivers and
D.~A.~Steer, ``Transport coefficients and analytic continuation in
dual 1+1
arXiv:hep-th/0204166.
}

\lref\BanadosWN{ M.~Banados, C.~Teitelboim and J.~Zanelli, ``The
Black Hole In Three-Dimensional Space-Time,'' Phys.\ Rev.\ Lett.\
{\bf 69}, 1849 (1992) [arXiv:hep-th/9204099];
M.~Banados, M.~Henneaux, C.~Teitelboim and J.~Zanelli, ``Geometry
of the (2+1) black hole,'' Phys.\ Rev.\ D {\bf 48}, 1506 (1993)
[arXiv:gr-qc/9302012].}

\lref\IsraelUR{ W.~Israel, ``Thermo Field Dynamics Of Black
Holes,'' Phys.\ Lett.\ A {\bf 57}, 107 (1976).
}

\lref\UnruhDB{ W.~G.~Unruh, ``Notes On Black Hole Evaporation,''
Phys.\ Rev.\ D {\bf 14}, 870 (1976).
}

\lref\ElitzurRT{ S.~Elitzur, A.~Giveon, D.~Kutasov and
E.~Rabinovici, ``From big bang to big crunch and beyond,'' JHEP
{\bf 0206}, 017 (2002) [arXiv:hep-th/0204189].
}

\lref\CrapsII{ B.~Craps, D.~Kutasov and G.~Rajesh, ``String
propagation in the presence of cosmological singularities,'' JHEP
{\bf 0206}, 053 (2002) [arXiv:hep-th/0205101].
}

\lref\CornalbaNV{ L.~Cornalba, M.~S.~Costa and C.~Kounnas, ``A
resolution of the cosmological singularity with orientifolds,''
arXiv:hep-th/0204261.
}

\lref\GubserBC{ S.~S.~Gubser, I.~R.~Klebanov and A.~M.~Polyakov,
``Gauge theory correlators from non-critical string theory,''
Phys.\ Lett.\ B {\bf 428}, 105 (1998) [arXiv:hep-th/9802109].
}

\lref\KeskiVakkuriNW{ E.~Keski-Vakkuri, ``Bulk and boundary
dynamics in BTZ black holes,'' Phys.\ Rev.\ D {\bf 59}, 104001
(1999) [arXiv:hep-th/9808037].
}

\lref\MaldacenaRE{ J.~M.~Maldacena, ``The large N limit of
superconformal field theories and supergravity,'' Adv.\ Theor.\
Math.\ Phys.\  {\bf 2}, 231 (1998) [Int.\ J.\ Theor.\ Phys.\  {\bf
38}, 1113 (1999)] [arXiv:hep-th/9711200].
}

\lref\LifschytzEB{ G.~Lifschytz and M.~Ortiz, ``Scalar Field
Quantization On The (2+1)-Dimensional Black Hole
Phys.\ Rev.\ D {\bf 49}, 1929 (1994) [arXiv:gr-qc/9310008].
}

\lref\BalasubramanianSN{ V.~Balasubramanian, P.~Kraus and
A.~E.~Lawrence, ``Bulk vs. boundary dynamics in anti-de Sitter
spacetime,'' Phys.\ Rev.\ D {\bf 59}, 046003 (1999)
[arXiv:hep-th/9805171].
}

\lref\WittenQJ{ E.~Witten, ``Anti-de Sitter space and
holography,'' Adv.\ Theor.\ Math.\ Phys.\  {\bf 2}, 253 (1998)
[arXiv:hep-th/9802150].
}

\lref\BalasubramanianRE{ V.~Balasubramanian and P.~Kraus, ``A
stress tensor for anti-de Sitter gravity,'' Commun.\ Math.\ Phys.\
{\bf 208}, 413 (1999) [arXiv:hep-th/9902121].
}

\lref\HorowitzXK{ G.~T.~Horowitz and D.~Marolf, ``A new approach
to string cosmology,'' JHEP {\bf 9807}, 014 (1998)
[arXiv:hep-th/9805207].
}

\lref\BalasubramanianDE{ V.~Balasubramanian, P.~Kraus,
A.~E.~Lawrence and S.~P.~Trivedi, ``Holographic probes of anti-de
Sitter space-times,'' Phys.\ Rev.\ D {\bf 59}, 104021 (1999)
[arXiv:hep-th/9808017].
}

\lref\CarneirodaCunhaNW{ B.~G.~Carneiro da Cunha,
``Three-dimensional de Sitter gravity and the correspondence,''
Phys.\ Rev.\ D {\bf 65}, 104025 (2002) [arXiv:hep-th/0110169].
}

\lref\MaldacenaKR{ J.~M.~Maldacena, ``Eternal black holes in
Anti-de-Sitter,'' arXiv:hep-th/0106112.
}

\lref\GiveonNS{ A.~Giveon, D.~Kutasov and N.~Seiberg, ``Comments
on string theory on AdS(3),'' Adv.\ Theor.\ Math.\ Phys.\  {\bf
2}, 733 (1998) [arXiv:hep-th/9806194].
}

\lref\TeschnerFT{ J.~Teschner, ``On structure constants and fusion
rules in the SL(2,C)/SU(2) WZNW
Nucl.\ Phys.\ B {\bf 546}, 390 (1999) [arXiv:hep-th/9712256];
``Operator product expansion and factorization in the H-3+ WZNW
Nucl.\ Phys.\ B {\bf 571}, 555 (2000) [arXiv:hep-th/9906215].
}

\lref\HorowitzJC{ G.~T.~Horowitz and D.~L.~Welch, ``Exact
three-dimensional black holes in string theory,'' Phys.\ Rev.\
Lett.\  {\bf 71}, 328 (1993) [arXiv:hep-th/9302126];
%
N.~Kaloper,
Phys.\ Rev.\ D {\bf 48}, 2598 (1993) [arXiv:hep-th/9303007];
%
M.~Natsuume and Y.~Satoh, ``String theory on three dimensional
black holes,'' Int.\ J.\ Mod.\ Phys.\ A {\bf 13}, 1229 (1998)
[arXiv:hep-th/9611041];
Y.~Satoh, ``Ghost-free and modular invariant spectra of a string
in SL(2,R) and
Nucl.\ Phys.\ B {\bf 513}, 213 (1998) [arXiv:hep-th/9705208];
%
J.~M.~Maldacena and A.~Strominger, ``AdS(3) black holes and a
stringy exclusion principle,'' JHEP {\bf 9812}, 005 (1998)
[arXiv:hep-th/9804085];
%
S.~Hemming and E.~Keski-Vakkuri,
Nucl.\ Phys.\ B {\bf 626}, 363 (2002) [arXiv:hep-th/0110252];
%
 J.~Troost,
``Winding strings and AdS(3) black holes,'' arXiv:hep-th/0206118;
%
E.~J.~Martinec and W.~McElgin, ``String theory on AdS orbifolds,''
JHEP {\bf 0204}, 029 (2002) [arXiv:hep-th/0106171];
}

\lref\BalasubramanianRY{ V.~Balasubramanian, S.~F.~Hassan,
E.~Keski-Vakkuri and A.~Naqvi, ``A space-time orbifold: A toy
model for a cosmological
arXiv:hep-th/0202187.
}

\lref\CornalbaFI{ L.~Cornalba and M.~S.~Costa, ``A New
Cosmological Scenario in String Theory,'' arXiv:hep-th/0203031.
}

\lref\NekrasovKF{ N.~Nekrasov, ``Milne universe, tachyons, and
quantum group'' arXiv: hep-th/0203112.
}

\lref\SimonMA{ J.~Simon, ``The geometry of null rotation
identifications,'' JHEP {\bf 0206}, 001 (2002)
[arXiv:hep-th/0203201].
}

\lref\LiuFT{ H.~Liu, G.~Moore and N.~Seiberg, ``Strings in a
time-dependent orbifold,'' JHEP {\bf 0206}, 045 (2002)
[arXiv:hep-th/0204168];
``Strings in time-dependent orbifolds,'' arXiv:hep-th/0206182.
}

\lref\LawrenceAJ{ A.~Lawrence, ``On the instability of 3D null
singularities,'' arXiv:hep-th/0205288.
}

\lref\FabingerKR{ M.~Fabinger and J.~McGreevy, ``On smooth
time-dependent orbifolds and null singularities,''
arXiv:hep-th/0206196.
}

\lref\HorowitzMW{ G.~T.~Horowitz and J.~Polchinski, ``Instability
of spacelike and null orbifold singularities,''
arXiv:hep-th/0206228.
}

\lref\SusskindQC{ L.~Susskind and J.~Uglum, ``String Physics and
Black Holes,'' Nucl.\ Phys.\ Proc.\ Suppl.\  {\bf 45BC}, 115
(1996) [arXiv:hep-th/9511227].
}

\lref\HartleAI{ J.~B.~Hartle and S.~W.~Hawking, ``Wave Function Of
The Universe,'' Phys.\ Rev.\ D {\bf 28}, 2960 (1983).
}

\lref\HartleTP{ J.~B.~Hartle and S.~W.~Hawking, ``Path Integral
Derivation Of Black Hole Radiance,'' Phys.\ Rev.\ D {\bf 13}, 2188
(1976).
}

\lref\NiemiNF{ A.~J.~Niemi and G.~W.~Semenoff, ``Finite
Temperature Quantum Field Theory In Minkowski Space,'' Annals
Phys.\  {\bf 152}, 105 (1984).
}

\lref\HemmingKD{ S.~Hemming, E.~Keski-Vakkuri and P.~Kraus,
``Strings in the extended BTZ spacetime,'' JHEP {\bf 0210}, 006
(2002) [arXiv:hep-th/0208003].
}

\lref\DixonJW{ L.~J.~Dixon, J.~A.~Harvey, C.~Vafa and E.~Witten,
``Strings On Orbifolds,'' Nucl.\ Phys.\ B {\bf 261}, 678 (1985).
}

\lref\StromingerCZ{ A.~Strominger, ``Massless black holes and
conifolds in string theory,'' Nucl.\ Phys.\ B {\bf 451}, 96 (1995)
[arXiv:hep-th/9504090].
}

\lref\JohnsonQT{ C.~V.~Johnson, A.~W.~Peet and J.~Polchinski,
``Gauge theory and the excision of repulson singularities,''
Phys.\ Rev.\ D {\bf 61}, 086001 (2000) [arXiv:hep-th/9911161].
}

\lref\BanksVH{ T.~Banks, W.~Fischler, S.~H.~Shenker and
L.~Susskind, ``M theory as a matrix model: A conjecture,'' Phys.\
Rev.\ D {\bf 55}, 5112 (1997) [arXiv:hep-th/9610043].
}

\lref\RastelliUV{ For a review of some recent developments see
L.~Rastelli, A.~Sen and B.~Zwiebach, ``Vacuum string field
theory,'' arXiv:hep-th/0106010.
}

\lref\WittenZW{ E.~Witten, ``Anti-de Sitter space, thermal phase
transition, and confinement in gauge theories,'' Adv.\ Theor.\
Math.\ Phys.\  {\bf 2}, 505 (1998) [arXiv:hep-th/9803131].
}

\lref\BalasubramanianZV{ V.~Balasubramanian and S.~F.~Ross,
``Holographic particle detection,'' Phys.\ Rev.\ D {\bf 61},
044007 (2000) [arXiv:hep-th/9906226].
}
\lref\HubenyDG{ V.~E.~Hubeny, ``Precursors see inside black
holes,'' arXiv:hep-th/0208047.
}

\lref\SusskindIF{ L.~Susskind, L.~Thorlacius and J.~Uglum, ``The
Stretched horizon and black hole complementarity,'' Phys.\ Rev.\ D
{\bf 48}, 3743 (1993) [arXiv:hep-th/9306069].
}

\lref\gks{N. Goheer, M. Kleban, and L. Susskind,  ``The Trouble
with de Sitter Space," [arXiv:hep-th/0212209].}

\lref\dks{L. Dyson, M. Kleban, and L. Susskind,  ``Disturbing
Implications of a Cosmological Constant," JHEP 0210:011,2002
[arXiv:hep-th/0208013].

\lref\MaldacenaBW{ J.~M.~Maldacena and A.~Strominger, ``AdS(3)
black holes and a stringy exclusion principle,'' JHEP {\bf 9812},
005 (1998) [arXiv:hep-th/9804085].
}

\lref\DysonNT{ L.~Dyson, J.~Lindesay and L.~Susskind, ``Is there
really a de Sitter/CFT duality,'' JHEP {\bf 0208}, 045 (2002)
[arXiv:hep-th/0202163].}
}

\lref\WittenQJ{ E.~Witten, ``Anti-de Sitter space and
holography,'' Adv.\ Theor.\ Math.\ Phys.\  {\bf 2}, 253 (1998)
[arXiv:hep-th/9802150].
}

\lref\AharonyTI{ O.~Aharony, S.~S.~Gubser, J.~M.~Maldacena,
H.~Ooguri and Y.~Oz, ``Large N field theories, string theory and
gravity,'' Phys.\ Rept.\  {\bf 323}, 183 (2000)
[arXiv:hep-th/9905111].
}
\lref\GiddingsPT{ S.~B.~Giddings and M.~Lippert, ``Precursors,
black holes, and a locality bound,'' Phys.\ Rev.\ D {\bf 65},
024006 (2002) [arXiv:hep-th/0103231].
}

\lref\FreivogelEX{ B.~Freivogel, S.~B.~Giddings and M.~Lippert,
``Toward a theory of precursors,'' Phys.\ Rev.\ D {\bf 66}, 106002
(2002) [arXiv:hep-th/0207083].
}

\lref\HorowitzFM{ G.~T.~Horowitz and V.~E.~Hubeny, ``CFT
description of small objects in AdS,'' JHEP {\bf 0010}, 027 (2000)
[arXiv:hep-th/0009051].
}

\lref\JacobsonMI{ T.~Jacobson, ``On the nature of black hole
entropy,'' arXiv:gr-qc/9908031.
}

\lref\BanksDD{ T.~Banks, M.~R.~Douglas, G.~T.~Horowitz and
E.~J.~Martinec, ``AdS dynamics from conformal field theory,''
arXiv:hep-th/9808016.
}

\lref\BalasubramanianDE{ V.~Balasubramanian, P.~Kraus,
A.~E.~Lawrence and S.~P.~Trivedi, ``Holographic probes of anti-de
Sitter space-times,'' Phys.\ Rev.\ D {\bf 59}, 104021 (1999)
[arXiv:hep-th/9808017].
}

\lref\KabatYQ{ D.~Kabat and G.~Lifschytz, ``Gauge theory origins
of supergravity causal structure,'' JHEP {\bf 9905}, 005 (1999)
[arXiv:hep-th/9902073].
}

\lref\DanielssonFA{ U.~H.~Danielsson, E.~Keski-Vakkuri and
M.~Kruczenski, ``Black hole formation in AdS and thermalization on
the boundary,'' JHEP {\bf 0002}, 039 (2000)
[arXiv:hep-th/9912209].
}

\lref\GregoryAN{ J.~P.~Gregory and S.~F.~Ross, ``Looking for event
horizons using UV/IR relations,'' Phys.\ Rev.\ D {\bf 63}, 104023
(2001) [arXiv:hep-th/0012135].
}

\lref\SusskindEY{ L.~Susskind and N.~Toumbas, ``Wilson loops as
precursors,'' Phys.\ Rev.\ D {\bf 61}, 044001 (2000)
[arXiv:hep-th/9909013].
}

\lref\HashimotoZP{ A.~Hashimoto, S.~Hirano and N.~Itzhaki, ``Large
branes in AdS and their field theory dual,'' JHEP {\bf 0008}, 051
(2000) [arXiv:hep-th/0008016].
}

\lref\HawkingDH{ S.~W.~Hawking and D.~N.~Page, ``Thermodynamics Of
Black Holes In Anti-De Sitter Space,'' Commun.\ Math.\ Phys.\ {\bf
87}, 577 (1983).
}

\lref\SusskindVU{ L.~Susskind, ``The World as a hologram,'' J.\
Math.\ Phys.\  {\bf 36}, 6377 (1995) [arXiv:hep-th/9409089].
}

\lref\GubserBC{ S.~S.~Gubser, I.~R.~Klebanov, and A.~M.~Polyakov,
``Gauge theory correlators from non-critical string theory,''
Phys.\ Lett.\ B {\bf 428}, 105 (1998) [arXiv:hep-th/9802109].
}

\lref\FidkowskiNF{ L.~Fidkowski, V.~Hubeny, M.~Kleban and
S.~Shenker, ``The black hole singularity in AdS/CFT,'' JHEP {\bf
0402}, 014 (2004) [arXiv:hep-th/0306170].
}

\lref\DrukkerZQ{ N.~Drukker, D.~J.~Gross and H.~Ooguri, ``Wilson
loops and minimal surfaces,'' Phys.\ Rev.\ D {\bf 60}, 125006
(1999) [arXiv:hep-th/9904191].
}

\lref\GrossGK{ D.~J.~Gross and H.~Ooguri, ``Aspects of large N
gauge theory dynamics as seen by string theory,'' Phys.\ Rev.\ D
{\bf 58}, 106002 (1998) [arXiv:hep-th/9805129].
}

\lref\CallanKZ{ C.~G.~.~Callan and J.~M.~Maldacena, ``Brane
dynamics from the Born-Infeld action,'' Nucl.\ Phys.\ B {\bf 513},
198 (1998) [arXiv:hep-th/9708147].
}

\lref\tHooftHY{ G.~'t Hooft, ``On The Phase Transition Towards
Permanent Quark Confinement,'' Nucl.\ Phys.\ B {\bf 138}, 1
(1978).
}

\lref\VafaTF{ C.~Vafa and E.~Witten,
Nucl.\ Phys.\ B {\bf 431}, 3 (1994) [arXiv:hep-th/9408074].
}

\lref\AharonyND{ O.~Aharony, Y.~E.~Antebi, M.~Berkooz and
R.~Fishman, ``'Holey sheets': Pfaffians and subdeterminants as
D-brane operators in large N gauge theories,'' JHEP {\bf 0212},
069 (2002) [arXiv:hep-th/0211152].
}

\lref\WittenXY{ E.~Witten, ``Baryons and branes in anti de Sitter
space,'' JHEP {\bf 9807}, 006 (1998) [arXiv:hep-th/9805112].
}

\lref\McGreevyCW{ J.~McGreevy, L.~Susskind and N.~Toumbas,
``Invasion of the giant gravitons from anti-de Sitter space,''
JHEP {\bf 0006}, 008 (2000) [arXiv:hep-th/0003075].
}

\lref\BalasubramanianNH{ V.~Balasubramanian, M.~Berkooz, A.~Naqvi
and M.~J.~Strassler, ``Giant gravitons in conformal field
theory,'' JHEP {\bf 0204}, 034 (2002) [arXiv:hep-th/0107119].
}

\lref\BalasubramanianZU{ V.~Balasubramanian and T.~S.~Levi,
``Beyond the veil: Inner horizon instability and holography,''
arXiv:hep-th/0405048.
}

\lref\AharonySX{ O.~Aharony, J.~Marsano, S.~Minwalla,
K.~Papadodimas and M.~Van Raamsdonk, ``The Hagedorn /
deconfinement phase transition in weakly coupled large N gauge
theories,'' arXiv:hep-th/0310285.
}

\lref\SusskindDR{ L.~Susskind, ``Matrix theory black holes and the
Gross Witten transition,'' arXiv:hep-th/9805115.
}

\lref\Minwallacom{S. Minwalla, et al., to appear}

\lref\ArnoldGH{ See for example P.~Arnold and L.~G.~Yaffe,
``Effective theories for real-time correlations in hot plasmas,''
Phys.\ Rev.\ D {\bf 57}, 1178 (1998) [arXiv:hep-ph/9709449].
}

\lref\SundborgUE{ B.~Sundborg, ``The Hagedorn transition,
deconfinement and N = 4 SYM theory,'' Nucl.\ Phys.\ B {\bf 573},
349 (2000) [arXiv:hep-th/9908001].
}

\Title{\vbox{\baselineskip12pt \hbox{hep-th/0406086}
\hbox{SU-ITP-04-24} \vskip-.4in}}  {D-Brane Instability as a Large
N Phase Transition}

\centerline{Lukasz Fidkowski and Stephen Shenker}
\bigskip

\centerline{\it Department of Physics, Stanford University,
Stanford, CA, 94305, USA}

\bigskip\bigskip \baselineskip14pt

\noindent In AdS/CFT analyticity suggests that certain singular
behaviors expected at large 't Hooft coupling should continue
smoothly to weak 't Hooft coupling where the gauge theory is
tractable. This may provide a window into stringy singularity
resolution and is a promising technique for studying the signature
of the black hole singularity discussed in hep-th/0306170.  We
comment briefly on its status. Our main goal, though, is to study
a simple example of this technique. Gross and Ooguri
(hep-th/9805129) have pointed out that the D-brane minimal surface
spanning a pair of 't Hooft loops undergoes a phase transition as
the distance between the loops is varied.  We find the analog of
this behavior in the weakly coupled Super Yang Mills theory by
computing 't Hooft loop expectation values there.

 \Date{June, 2004}

\newsec{Introduction}

\ifig\figGeodesics{The Penrose diagram for the AdS-Schwarzschild
black hole in $d=5$.  We have drawn the three geodesics connecting
points with $t=0$ on the two asymptotic boundaries (at finite AdS
cutoff), and the null $t_c$ geodesic.} {\epsfxsize=7cm
\epsfysize=7cm \epsfbox{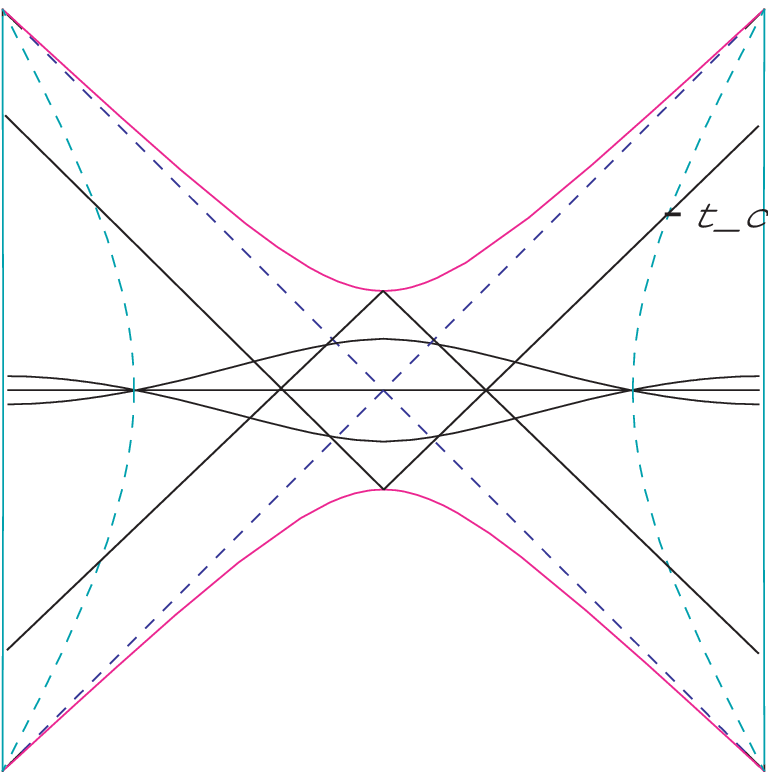}}

In \FidkowskiNF\  we and our collaborators used the AdS/CFT
correspondence to identify a subtle but distinct signature of the
black hole singularity in a boundary thermal gauge theory
correlator. This work built on earlier insights into studying
physics behind the horizon using boundary correlators in AdS/CFT
\refs{\BalasubramanianZV, \LoukoTP, \MaldacenaKR, \KrausIV,
\LeviCX }.\foot{For subsequent work see \refs{
\BalasubramanianZU}.}  In \FidkowskiNF\ we studied boundary
operators $\CO $ in D=4 SYM that create D3-branes in the AdS$_5
\times S^5$ bulk wrapped on the $S^5$ \refs{\WittenXY,
\McGreevyCW, \BalasubramanianNH}. These behave like pointlike
particles in the AdS$_5$ space. The mass $m$ of these particles is
$\sim N$ and so in the large $N$ limit becomes infinite. In the
bulk, correlators of these operators can be computed precisely in
the supergravity limit by using the geodesic approximation,
$\langle\CO(x) \CO(y)\rangle \sim \exp(-m \len)$ where $\len$ is
the proper length of the geodesic connecting $x$ and $y$.   In
\FidkowskiNF\ we examined correlators of two such operators,  one
placed on each of the two disjoint boundaries of the eternal
AdS-Schwarzschild black hole spacetime \MaldacenaKR\
(\figGeodesics).  Even though these operators are spacelike
separated their expectation value is bulk diffeomorphism
invariant. This is different than the situation in asymptotically
flat spaces where the only available diffeomorphism invariant
quantities  are S-matrix elements,  which involve timelike or null
separated sources. This subtlety of asymptotically AdS spaces
opens up an important window. Geodesics between these points
(\figGeodesics) pass behind the black hole horizon and so their
behavior yields information about the geometry there.

\ifig\figAnalyticStructure{The analytic structure of ${\cal C}(t)$
in the complex $t$ plane.  The horizontal axis corresponds to
Minkowski time, the vertical axis to Euclidean . There is a
periodic identification $t=t+i \beta$ in the $t$ plane. $t = i
{\beta \over 2}$ corresponds to coincident points and the branch
cut there (not shown) reflects the usual coincident points pole in
$C(t)$. Near $0$, ${\cal C}(t)$ has a $t^{4 \over 3}$ branch cut
and the singularity corresponding to the null geodesic is at $t_c$
on the second sheet.  One gets to it by analytically continuing
around the branch cut by $180$ degrees, as indicated.}
{\epsfxsize=7cm \epsfysize=5cm \epsfbox{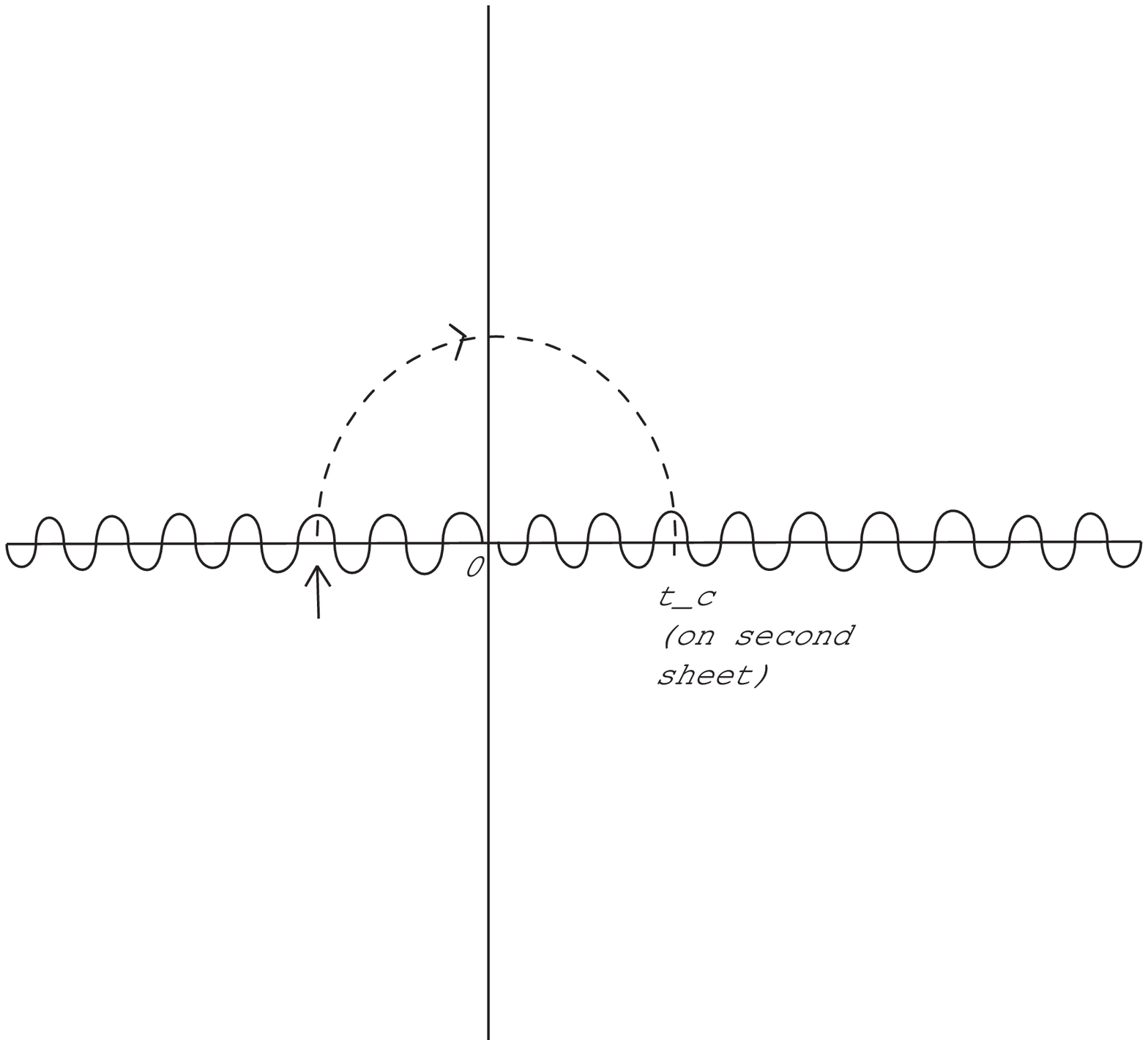}}

In fact it is possible to ``move" an operator from one asymptotic
boundary to another by analytically continuing in complexified
boundary time $t$.   All correlators are periodic with period $i
\beta$ where $\beta$ is the inverse temperature of the black hole.
Moving from one boundary to another involves shifting
$t\rightarrow -t+ i \beta/2$.  So the correlator we are interested
in can be represented as a thermal correlator $C(t) = \langle
\CO(t)\CO(-t+i\beta/2)\rangle$.  In fact, it will be more
convenient to work with ${\cal C}(t) = \log C(t) / m$.   The
analytic structure of ${\cal C}(t)$ determined in \FidkowskiNF\ is
summarized in \figAnalyticStructure. There is a cube root branch
cut at $t=0$  that results from the coincidence of three geodesics
linking the boundary points.  Two of these geodesics ``annihilate"
and then become complex, much like roots of an analytic equation.
On the second sheet, the correlator becomes dominated by a
geodesic that at a certain $t=t_c$ ``bounces" off the black hole
singularity when the geodesic is nearly null (\figGeodesics). The
vanishing of its proper length means ${\cal C}(t) \sim - 2 \log
(t-t_c)$.  This $t_c$ singularity is a direct consequence of the
diverging curvature at the black hole singularity and so provides
a distinct signature of it in the boundary theory.   The analysis
described above is done in the supergravity limit, $g_s , l_s
\rightarrow 0$ or $\lambda, N \rightarrow \infty$, where $l_s$ is
the string length and $ \lambda$ is the 't Hooft coupling.

The boundary gauge theory defined conventionally would actually
yield results described by the first sheet in
\figAnalyticStructure. Here the bulk correlator is dominated for
general $t$ by complexified geodesics.   But knowledge of these
correlators on the first sheet is enough to determine the the
behavior of ${\cal C}(t)$ on the second sheet.   A modest number
of terms in a Taylor expansion of ${\cal C}(t)$ allows one to
determine $t_c$ and the strength of the singularity there with
high accuracy by standard extrapolation techniques. The
singularity at $t_c$ is a bit like a broad resonance and so its
presence is reflected in a distinct but broad feature on the first
sheet.\foot{This technique depends on the analyticity of various
quantities. One might argue that we should just take the metric,
computed outside the horizon, and analytically continue it to the
singularity. This would be a reasonable thing to do except for the
fact that in string theory the local metric is not meaningful. The
virtue of the present approach is that we have located a signature
of the black hole singularity in a system where all quantities are
nonperturbatively well defined and analytic.}

In principle we could determine the effects of small but finite
$l_s$ ($\alphaprime$) on the black hole singularity by obtaining
boundary gauge theory data at finite but large $\lambda$ and
extrapolating to the $t_c$ singularity.\foot{Finite $g_s$ is more
subtle because the mass $m \sim \lambda/g_s$ must be held larger
than all other quantities.  Certain quantities, like power series
in $g_s$, leading nonperturbative effects, and certain scaling
limits should be available, though \FidkowskiNF.} But we encounter
the standard problem in employing dualities.   The gauge theory at
large $\lambda$ is not effectively computable by gauge theory
techniques.

But here we can take advantage of the severity of the problem we
are studying.   The black hole singularity is sufficiently
mysterious that we would be happy to learn even something
qualitative about its behavior. We expect the large $N$  classical
behavior of the gauge theory to be analytic in $\lambda$ except at
$\lambda = \infty$.  The small $\lambda$ expansion has a finite
radius of convergence since there are only an exponential number
of planar Feynman diagrams at each order. As $\lambda \rightarrow
\infty$ the tree level string dual has an asymptotic expansion in
$\alphaprime \sim 1/\lambda^{1/2}$.  World sheet instantons
produce effects of  the form  $\exp(-1/\alphaprime) \sim
\exp(-\lambda^{1/2})$ and  we expect an essential singularity at
$\lambda = \infty$.  The analyticity tells us that if the $t_c$
singularity persists to finite large $\lambda$ (at $g_s =0$) then
it must evolve analytically all the way to $\lambda =0$. It can
collide with other singularities and move off into the complex
plane but it cannot just vanish.  So the qualitative aspects of
the singularity can be examined at small $\lambda$. Here weak
coupling Feynman diagram techniques are effective.   The presence
of an essential singularity  at $\lambda = \infty$ allows the
$t_c$ singularity to vanish abruptly there, and be absent for all
finite $\lambda$.  This can be diagnosed at weak coupling.

This procedure of following large $\lambda$ singularities to the
small $\lambda$ limit has been employed in the study of the
Hawking-Page transition to the black hole \refs{\WittenZW,
\SundborgUE \AharonySX}.  In the gauge theory this is just a large
$N$ ``deconfinement" transition. The Gregory-Laflamme transition
has been described in a similar way \refs{\SusskindDR,
\Minwallacom}.

The situation we are studying here is somewhat different because
we are in the high temperature, ``large black hole", $N^2$ entropy
phase of the gauge theory for all $\lambda$. There is no bulk
phase transition.  Instead we  are looking at the dynamics of a
single D-brane. This will involve $N$ rather than $N^2$ degrees of
freedom.

The most straightforward singularity to study is not the $t_c$
singularity but the branch cut at $t=0$ involving the coalescence
of D-brane geodesics (\figGeodesics).

In this paper we study a ``practice problem" where there is a
coalescence of D-brane saddle points involving only $N$ degrees of
freedom and track it to small $\lambda$.

\ifig\figthefirst{(a) For $s < s_0$, the connected catenoid
dominates. (b) For $s > s_0$, the disconnected disc solution
dominates.} {\epsfxsize=7cm \epsfysize=7cm \epsfbox{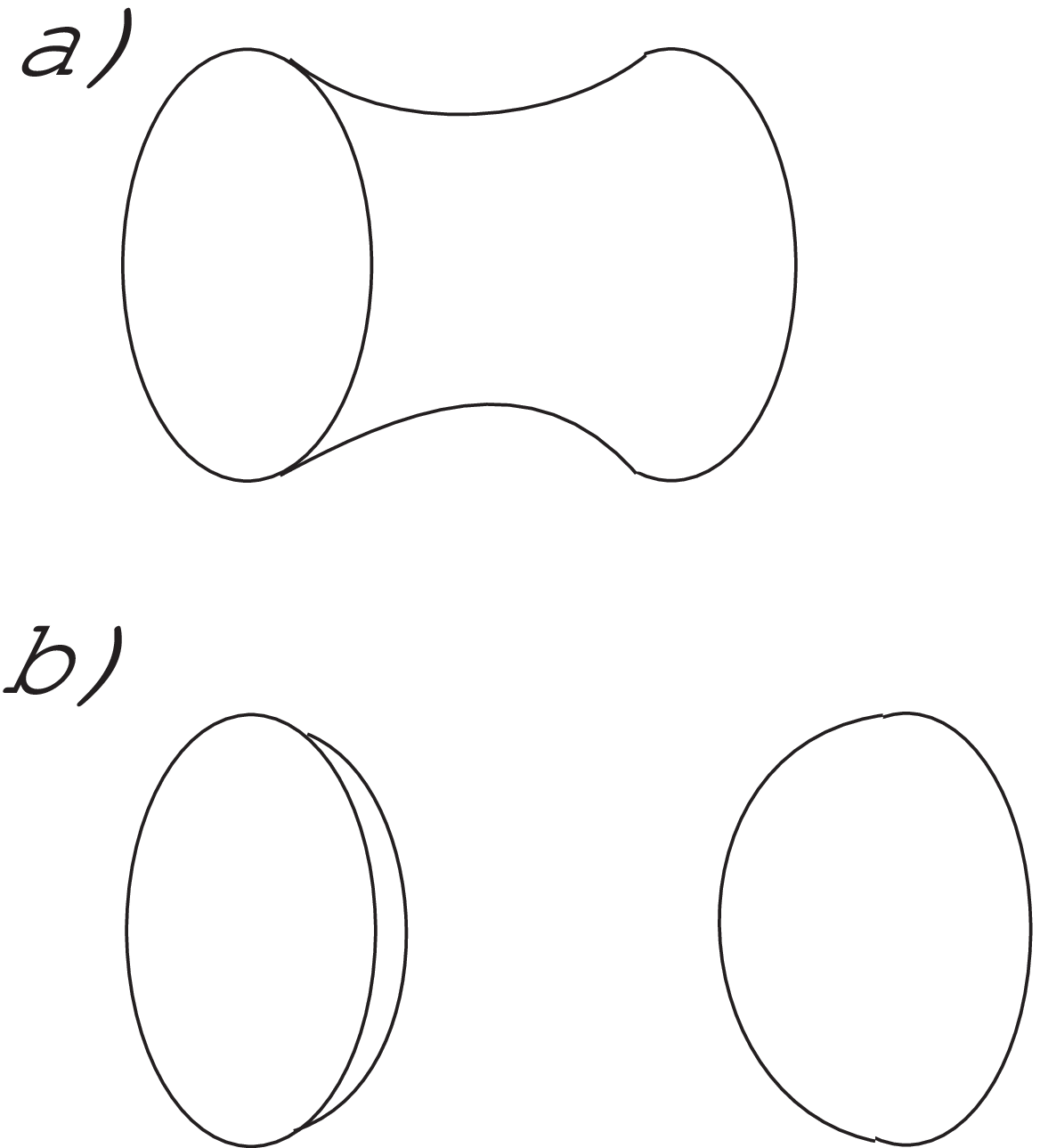}}

The problem we study was first described by Gross and Ooguri
\refs{ \GrossGK, \DrukkerZQ}.  It involves the correlator of two
't Hooft loops in the boundary gauge theory at zero temperature.
In the bulk this correlator is determined by the world sheets of
D-strings that end on the 't Hooft loop. The tension of the
D-string world sheet is $l_s^2/g_s \sim N/\lambda^{1/2}$.   At
infinite $N$ the tension is infinite, independent of $\lambda$, so
we can use saddle point techniques to pick out the dominant world
sheet.\foot{Gross and Ooguri also discuss  the Wilson loop
correlator. At any finite $\lambda$ the fundamental string tension
is finite so the transitions discussed below are smoothed out.}
Consider circular loops of radius $R$ oriented face to face with
separation $L$.  By conformal invariance things will depend only
on the ratio $s = L/R$. For $s \ll 1$ the dominant saddle is a
catenoid connecting the loops. At $s=s_0$ another saddle becomes
dominant: two disconnected discs one terminating on each loop (see
\figthefirst). This ``phase transition" is first order. The
catenoid continues to exist for $s> s_0$, analogous to a
metastable phase. Its properties can be studied by analytic
continuation from the $s < s_0$ regime.  At $s=s_1 > s_0$ the
catenoid solution ceases to exist.  Another catenoid, a local
maximum of the action, coalesces with the stable catenoid and they
both go off into the complex plane, analogous to a second order
phase transition.  The radius of curvature of these surfaces is of
order $L$ at the phase transitions, which can be taken much
greater than the AdS radius which in turn is much greater than the
string length at large $\lambda$.  So we do not expect $\alpha'$
effects to destabilize these transitions at large but finite
$\lambda$. So we expect them to move analytically all the way to
small $\lambda$.

Our main calculational goal in this paper is to study these phase
transitions at weak gauge coupling.  More precisely we will study
the  correlator of two 't Hooft loops at $N=\infty$ and small
$\lambda$ where semiclassical techniques in the gauge theory are
applicable.  We find that at small $\lambda$ the two phase
transitions have merged into a second order transition occurring
at a critical separation $s_0 (\lambda)$.  The small $\lambda$
large $N$ Feynman diagram analysis also suggests a connection
between the second order transition and the strong coupling
D-brane picture.

\newsec{Construction of the 't Hooft Loop}

Let us introduce some conventions.  We will be working with the
Euclidean Ad$S^5 \times S^5$ metric: \eqn\adsmetric{ ds^2 = {R^2
\over Y^2} \( \sum_{\mu=0}^3 dX^\mu dX^\mu + \sum_{i=1}^6 dY^i
dY^i \)} with the AdS radius $R$ given by $R^4 = \lambda
{\alpha'}^2$ and $Y^2 = \sum_{i=1}^6 {Y_i}^2$.  The dual gauge
theory is ${\cal N} = 4$ SYM, which in addition to the gauge
fields has six scalars $\Phi_i$, $i = 1, \cdots, 6$, all in the
adjoint.  The bosonic part of the action is \eqn\bosonicaction{S =
{1 \over 2 g^2} \int d^4 x Tr \left({1 \over 2} F_{\mu \nu} F^{\mu
\nu} + \sum_{i=1}^6 (D_\mu \Phi_i)^2 + \sum_{1 \leq i < j \leq 6}
[\Phi_i, \Phi_j]^2\right)}  The gauge theory supersymmetric Wilson
loop operator \eqn\wilsonloop{ W({\cal C}) = \Tr \left[ P \exp
\int_{\cal C} \left( i {A_\mu} {\dot x}^\mu + \theta^I \Phi^i
\sqrt{{\dot x}^2} \right) d \tau \right] } can be evaluated in the
bulk by integrating over fundamental string worldsheets attached
to the the Wilson loop, weighted by $\exp (-S)$ where $S$ is the
Polyakov action \AharonyTI.  As discussed above, the tension of
the fundamental string world sheet is finite for finite $\lambda$
so all phase transitions are smeared out.  For this reason we
focus on 't Hooft loops, which are the electric-magnetic duals of
Wilson loops and are evaluated in the bulk using D-string
worldsheets, whose tension is proportional to $N$, and hence can
be taken infinite independent of $\lambda$. We can write out the
gauge theory operator corresponding to a 't Hooft loop just by
replacing the gauge field with its electric magnetic dual.  To
actually evaluate it (at small $\lambda$) we follow the original
construction.






   The 't Hooft loop was originally defined \tHooftHY\ in
an arbitrary Yang-Mills theory all of whose fields are invariant
under the center $Z_N$ of $SU(N)$. Before we discuss the loop,
let's consider the simpler case of the 't Hooft vortex in 2 + 1
dimensional Yang Mills, also defined in \tHooftHY. The vortex can
be thought of as a monopole in 3 dimensions. More precisely, we
define the vortex operator $\phi(x,t)$ in the Minkowski theory as
an operator which acts on each field configuration by a
multivalued gauge rotation whose holonomy around the point $x$ is
$e^{2 \pi i \over N}$, which is in the center $Z_N$.  Then the
computation of the correlator of these operators amounts to a path
integral over field configurations with Dirac strings running
between the vortex operators. For example, the two point function
of $\phi$ and $\phi^*$ is given by \eqn\vortextwopoint{ (
\phi(0,t) \phi^*(0,0) ) = {\int_C e^{-S} \over \int e^{-S}}} where
the integral in the numerator is over all field configurations
with a Dirac string running from one vortex to the other.  The
holonomy of the gauge field around the Dirac string is $e^{2 \pi i
\over N}$.  Thus in the saddle point approximation the logarithm
of the two point function is the energy of two monopoles with
opposite charges separated by a distance $t$ in $3$ dimensions.

The 't Hooft loop is the analog in $3+1$ dimensions of the 't
Hooft vortex, and the expectation value of a product of such loops
is computed by doing a path integral over all field configurations
with appropriate $2$ dimensional Dirac sheets connecting the
loops.

In order to evaluate the 't Hooft loop, we are instructed to do a
path integral with certain boundary conditions at the Dirac
worldsheet.  Now, as 't Hooft \tHooftHY\ argues, the lowest action
configurations will be ones in which the gauge field is that of a
monopole whose world line is the loop and which lies in some
$U(1)$ that is conjugate in the $SU(N)$ Lie algebra to the $U(1)$
of the form diag $(N - 1, -1, \cdots, -1)$.  This is consistent
with the holonomy condition.  After we normalize and take the
large $N$ limit, these $U(1)$'s are basically all $SU(N)$
conjugates of diag $(1, 0, \cdots, 0)$. This set of matrices is
just $SU(N) / (U(1) \times SU(N-1)) = {CP}^{N-1}$. We denote this
space $M$.  It will be very useful to us in the rest of this
paper.  We note that we can parametrize it by matrices of the form
${\cal M}_{ij} = (u_i {u_j}^*)$ where $\Sigma u_i {u_i}^* = 1$.
The choice of $u_i$ is unique except for an overall phase, and the
parametrization in fact defines an $S^1$ fibration $S^{2N-1}
\rightarrow {CP}^{N-1}$.  Also note that while $M$ captures all of
the possible boundary conditions for the 't Hooft loop, its
adjoint has boundary conditions which can lie in any $U(1)$
conjugate to diag $(-1, 0, \cdots, 0)$.  These boundary conditions
are entirely disjoint from the original $M$, i.e. the set of
$SU(N)$ conjugates of diag $(1, 0, \cdots, 0)$ and diag $(-1, 0,
\cdots, 0)$ comprise two disjoint copies of $M$.  We refer to the
adjoint boundary conditions as having negative gauge charge.

The loop also sources a scalar field, which one might think can be
in any $U(1)$. However, if the state is not BPS then even though
it has the same tree level energy as a BPS state, it gets infinite
radiative corrections as the cutoff (i.e. the "thickness" of the
monopole) is taken to zero.
 Thus we
only need to consider BPS states. Setting the variation of the
gaugino to zero, we have that the condition for unbroken
supersymmetry is $\Gamma^{\mu \nu} F_{\mu \nu} \epsilon = 0$,
where $\epsilon$ is a Dirac spinor. The classical solutions for a
straight line are (\CallanKZ) \eqn\classicalmonopole{
F_{\hat{\theta} \hat{\phi}} = X^9 = {\pi N \over r}}
\eqn\classicalmonopoleb{ F_{\hat{\theta} \hat{\phi}} = -X^9 = {\pi
N \over r}} The BPS conditions reduce to
\eqn\bpscondition{(\Gamma^{\hat{\theta} \hat{\phi}} + \Gamma^{r
9}) \epsilon = 0} \eqn\bpsconditionb{(\Gamma^{\hat{\theta}
\hat{\phi}} - \Gamma^{r 9}) \epsilon = 0} and are satisfied by
half of the spinors $\epsilon$ as one can easily check.  We see in
particular that the gauge field and scalar will always be in the
same $U(1)$.

We would like to have some information about the solution for a
circular 't Hooft loop.  One could make general arguments to
obtain this information, but we will just do an explicit conformal
transformation to make the straight line into a circle of radius
$R$ that sits centered at the origin of the $(x,t)$ plane.  One
finds that the necessary transformation is
\eqn\conformaltransform{ (t,x,y,z) \rightarrow {2 R^2 \over u^2 +
R^2 + 2xR} (t, {u^2 - R^2 \over 2R}, y, z)} where $u^2 = t^2 + x^2
+ y^2 + z^2$.  The $U(1)$ solution for the straight line up to
sign and numerical factors is $F_{ij} = ({x_i}^2)^{-{3 \over 2}}
\epsilon_{ijk} x_k dx_i dx_j$ and $X_9 = ({x_i}^2)^{-{1 \over
2}}$, where we have introduced the spatial indices $i,j,k = 1, 2,
3$.  The image of this solution under the conformal transform is
\eqn\transformedfield{ F = {4 R^2 \over (t^2 + x^2 + y^2 + z^2 +
R^2)^3} \times \omega} \eqn\transformedfieldcont{ \omega = (2xz)
dx \wedge dy + (-t^2 - x^2 + y^2 + z^2 + R^2) dy \wedge dz + (2xy)
dz \wedge dx + (2zt) dy \wedge dt - (2yt) dz \wedge dt}
\eqn\transformedfieldagain {X^9 = {2R \over t^2 + x^2 + y^2 + z^2
+ R^2}} Notice that the field strength decreases as the fourth
power of the distance, and the scalar as the square of the
distance.

\newsec{Two 't Hooft Loops}

Now that we have these preliminaries out of the way, we can look
at the correlator $H(s) = - {1 \over N} \log (T_1 {T_2}^*)$ at
small $\lambda$, where $s = L/R$ as defined above. For
concreteness we will work in a fixed gauge, e.g.,  $\del^\mu A_\mu
= 0$.

 As discussed above, we have a certain
classical configuration corresponding to $T_1$; the one
corresponding to the adjoint ${T_2}^*$ has negative gauge charge.
It also has the same scalar charge.  We will make the ansatz that
the lowest action configuration for two loops has their boundary
conditions in the same $U(1)$ and is the sum of the individual
loop solutions (\tHooftHY). Note that it is not strange that the
two scalar charges should be the same, since like scalar charges
attract.


The configurations that contribute to the path integral have
boundary conditions at each loop, so we can divide the
configurations up into sectors parametrized by $M \times M$, where
we think of the two $M$'s as parametrizing the boundary conditions
at the two loops.  This way of parametrizing configurations is not
canonical, since it makes use of the fact that we are working in a
particular gauge, but will nevertheless be useful.  For each point
in $M \times M$, i.e. each choice of boundary conditions, there is
a lowest action field configuration with those boundary
conditions.  When the boundary conditions are the same, the
minimal configuration is just the sum of the individual 't Hooft
loop solutions, as described above.  When they are orthogonal,
i.e. if $\Tr (AB) = 0$ for $(A,B)$ in $M \times M$, the sum of the
individual one 't Hooft loop solutions in those $U(1)$'s gives a
minimum for the action.  Note that in this case the action is
independent of the separation between the loops.  We will refer to
the preceding two possibilities as "same $U(1)$'s" and "orthogonal
$U(1)$'s" respectively.

We want to define a parameter which measures the relative
orientation of the two $U(1)$'s.  $y = \Tr (AB)$ for $(A,B)$ in $M
\times M$ would seem to be a natural choice.  Letting $B = $ diag
$(1, 0, \cdots, 0)$ for simplicity we get $y = u_1 {u_1}^*$ in
terms of the previously defined parametrization of ${CP}^{N-1}$.
Thus $y$ ranges from $1$ to $0$, with $y=1$ for same $U(1)$'s and
$y=0$ for orthogonal $U(1)$'s; in particular, note that $y$ is
always positive.  However, $y$ is not a good coordinate on $M
\times M$ near orthogonal $U(1)$'s because it is easy to see that
for any smooth path on $M$ parametrized by some coordinate $\tau$
and starting at a $U(1)$ orthogonal to diag $(1, 0, \cdots, 0)$,
we have $y(\tau)$ going like at least $\tau^2$ for small $\tau$.
In other words, the appropriate coordinate near orthogonal
$U(1)$'s is in a sense the square root of $y$, so we define $u = y
^ {1 \over 2}$.


\ifig\figEntropy{A graph of the energy ${\cal E}$ (lower line) and
minus the entropy $- {\cal S}$ (upper line) as a function of $u$
on $M$. The energy is maximized at the point of largest entropy. }
{\epsfxsize=7cm \epsfysize=5cm \epsfbox{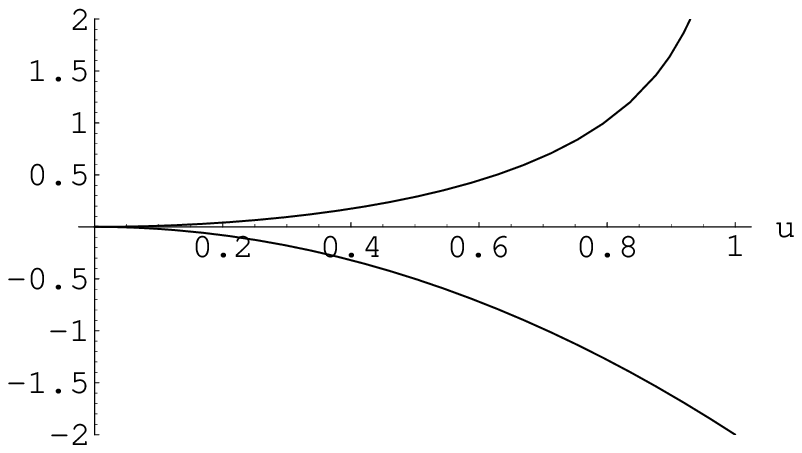}}

We now give an intuitive description of the weak coupling behavior
of $H(s)$.   For small $s$ it is favorable for the loops to take
advantage of the attractive magnetic interaction energy between
them and orient their monopoles in the same $U(1)$.  That is $u
\sim 1$.   But as $s$ increases the interaction energy decreases
and the loops start taking advantage of the very large number of
near orthogonal $U(1)$ directions.  Entropy begins to dominate.
Because $N$ is infinite this transition is sharp. After the
transition the two $U(1)$'s are basically uncorrelated and so we
would expect that $H(s)$ is independent of $s$. This is
reminiscent of the strong coupling picture, where after the
transition to the disconnected caps $H(s)$  becomes independent of
$s$ as well.  We will discuss this connection in more detail
below.

We now make our intuition precise.  We define an effective action
functional on $M \times M$ by integrating over all field
configurations with specified boundary conditions:
\eqn\effectiveaction{\exp (-N f_{AB}) = \int_{AB} \exp (-S)} Here
$\int_{AB}$ is the path integral over configurations with boundary
conditions $A$ and $B$ at the first and second 't Hooft loop
respectively.  Since $f_{AB}$ depends only on $u$, we will use the
notation $f(u)$. The tree level contribution to $f_{AB}$ (denoted
${\cal E}$) is just the action of the classical field
configuration with those boundary conditions.  For the regime we
are interested in, namely large separation $s$ and small coupling
$\lambda = g^2 N$, the important, $s$ dependent part of the action
comes from points that are much farther away than $R$ from either
loop, and there the spacetime derivative term in the field
strength dominates the commutator and so the sum is approximately
a solution of the equations of motion. Its action is just
proportional to $u^2= \Tr AB $.  Higher order terms in $u$ will
come from corrections to the approximation we use and radiative
corrections, suppressed by the coupling or the inverse separation.
One can get the dependance on $s$ using the $r^{-2}$ fall off of
the scalar at large distances, and obtain \eqn\energy{ {\cal E}= -
{N  \over \lambda} c u^2 s^{-2}} where $c$ is a numerical constant
of order one.

In addition to the tree level contribution, there are clearly many
gauge field configurations with the same $u$ so there is a large
collective coordinate zero mode integral to do.   (There are
further nonzero mode radiative corrections but they are small at
small $\lambda$. We discuss them later.)  The integral over the
zero modes (which can be associated with global gauge
transformations) produces the entropy discussed above.  The
effective action functional with values of $u$ between $u$ and
$u+du$ gets weighted by an extra factor of $\exp ({\cal S}(u))
du$, equal to the volume of $M \times M$ between $u$ and $u + du$.
Here the volume is defined with respect to the unique metric on $M
\times M$ that is of product form and invariant under the $SU(N)$
symmetry.  In the large $N$ limit \eqn\entropy{ {\cal S}(u) = {1
\over 2} N \log (1-u^2).} This is just the entropy discussed
above.   The effective action at weak coupling (up to radiative
corrections) is then just \eqn\freeenergy{N f(u)= {\cal E} - {\cal
S} = N( -{1 \over \lambda}c u^2 s^{-2} - {1 \over 2} \log
(1-u^2))} The value of $u$ that minimizes \freeenergy\ is favored.
At large $N$ there are no fluctuations around this minimum.

\ifig\figFreeEnergy{A graph of the free energy $f(u)$ for several
values of the separation $s$.  Note that for large $s$ the value
of $u$ that minimizes $F$ is $0$.  We have  doubled the domain of
$u$ for clarity.  The graph reflects the true nature of the
transition near orthogonal $U(1)$'s on $M \times M$.}
{\epsfxsize=7cm \epsfysize=7cm \epsfbox{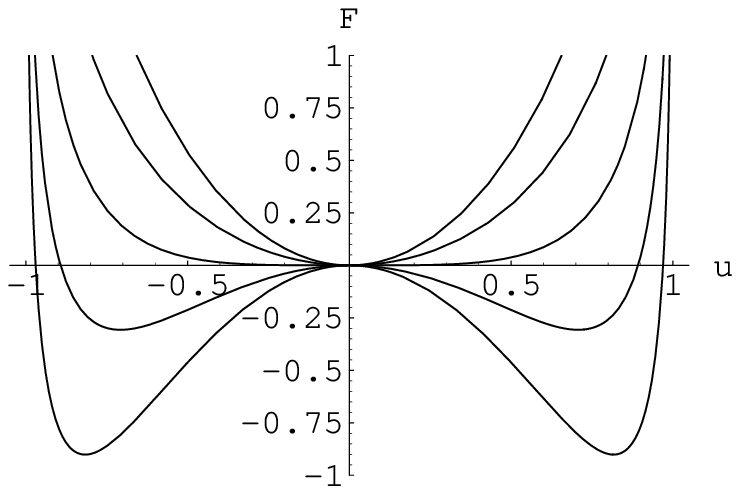}}

\ifig\figMiny{A graph of the value of $u$ that minimizes the free
energy $F$, as a function of the separation $s$.} {\epsfxsize=7cm
\epsfysize=5cm \epsfbox{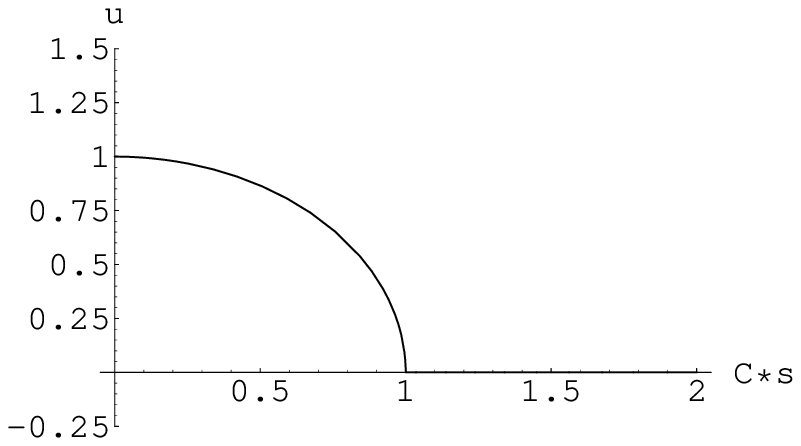}}

In \figFreeEnergy\  we show how the profile of $f$ changes as we
vary $s$ and in \figMiny\ the resulting profile of the minimum
$u$.   We see that this a standard second order phase transition
with the coefficient of a quadratic minimum vanishing at some
$s=s_0 (\lambda)$.  When $s$ is near $s_0 (\lambda)$
\eqn\quadraticvan{f(u) = a_1(s-s_0 (\lambda)) u^2 + a_2 u^4 +
\cdots}  and because the expansion of $-\log(1-u^2)$ contains all
{\it positive} order $1$ coefficients, all the $a_i$ are positive.
Note that radiative corrections will have small coefficients
(suppressed by $\lambda$ and/or powers of the inverse separation)
and so cannot change the nature of the transition. Thus indeed the
transition is second order (i.e. the $u$ that minimizes $f(u)$ is
a continuous function of $s$).  It follows that  $H(s)$ (the log
of the 't Hooft loop correlator)  goes like $(s-s_0(\lambda))^2$
for $s < s_0 (\lambda)$ and is $0$ for $s > s_0 (\lambda)$.  In
fact, $H(s)$ is {\it identically} $0$, even including all
radiative corrections.  We explain this at the end of this
section.  To begin, we discuss general radiative corrections to
\freeenergy.

First we argue qualitatively.  The effective action expanded
around  a classical configuration  like those discussed above will
have the following schematic form \eqn\logdet{ S_{eff} = N \int
d^4 x  [A \Tr F^2 + B \Tr (DF)^2 + C \Tr F^4 + \cdots]} where $F$
is the classical gauge field strength.  We have omitted scalars
for simplicity.  The coefficient $A$ vanishes because the beta
function vanishes.   The coefficents $B$ and $C$ are IR divergent.
The natural IR cutoff is $L$, the separation between loops, so $B
\sim L^2$ and $C \sim L^4$.  $F$ goes like $1/L^3$ so the second
term is $\sim 1/L^6$ and the third term is $\sim 1/L^8$.   Both of
these terms (and all others) are small compared to the leading
classical term $\sim {1 \over \lambda L^6}$.

We will now give an argument that the coefficients in front of all
terms in the effective action are finite to one loop.  We find the
fluctuation determinant using the background field method,
expanding about one of the classical solutions with some boundary
conditions $A,B$ at the two loops. The idea is to write the log of
the determinant as in \logdet\
 and find the coefficients in front of
all the terms using the standard 1PI Feynman diagram expansion. In
fact we have separate determinants for the gauge field, scalar,
and fermions. However, we can manipulate them to all look like
determinants associated to Laplacians in some background (for the
fermions we have to look at the square of the determinant), so
that the final prescription becomes to sum over one loop diagrams
with scalar propagators in the loop and background field
insertions along it. Each momentum injected in an insertion is
weighted by a factor equal to the fourier transform of the
background field at that momentum, so that for each diagram we
have to integrate over the loop momentum and over the external
momenta.  Let us consider a diagram with $n$ insertions; it will
be convenient to parametrize the momentum integrals by $p_i, i =
1, \cdots, n$, where $p_i$ is the momentum in the propagator
between insertion $i$ and $i+1$. The diagram is then
\eqn\feynmandiagram{ \int \left(\prod_i d^4 p_i \right) \left(
\prod_i {1 \over {p_i}^2} \right) \left( \prod_i B_i(p_i -
p_{i+1}) \right) } where $B$ is the fourier transform of one of
the background fields. Let us look at potential IR divergences
first.  For small $p$ we find, using the explicit one t' Hooft
loop solution above, that $B(p)$ goes like ${1 / p^2}$ for the
scalar and like $\log(p)$ for the gauge field strength.  Only
diagrams with all scalar insertions (and no scalar derivatives)
could conceivably have an IR divergence and by general arguments
these would not contribute to the $s$-dependent part of the
effective action.  In this particular theory we also know these
vanish because of the existence of flat directions. The other
diagrams are superficially IR convergent and in fact one can argue
that they indeed are IR convergent.  In the UV $B(p)$ goes like
${1 / p^2}$ for both the gauge and scalar fields so the diagram is
logarithmically UV divergent.  However, because the two 't Hooft
loop state is BPS on scales smaller than $L$ this logarithmic
divergence is cancelled and the diagram is superficially UV
finite; again, one can argue that it actually is UV finite. We
should note that potential $UV$ divergences can come from
integrations over the external momenta in addition to coming from
the loop integral, and all diagrams have to be checked. Indeed,
even the tree diagram corresponding to the classical action is UV
divergent, reflecting the fact that the classical action of our
configuration is infinite.  The usual loop UV divergence, occuring
in diagrams corresponding to $F^2$ and $(\del X)^2$, however, is
cancelled just because of supersymmetry and does not rely on the
BPS nature of the state.

\ifig\figOrtPert{(a) When the boundary conditions are not
orthogonal, we can have type I and type II insertions from the
same $U(1)$ (among other possibilities).  The resulting (leading
in $1/N$) radiative corrections correspond to disk worldsheets
capping the connected $U(1)$'s configuration. (b) With orthogonal
boundary conditions it is impossible to have radiative corrections
that correspond to a worldsheet with disk topology. (c) Instead,
the radiative corrections are down by a factor of $N$ and
correspond to worldsheets with cylindrical or higher genus
topology connecting the two D-branes.  There can be no disk
worldheets connecting two disconnected D-branes. } {\epsfxsize=8cm
\epsfysize=7cm \epsfbox{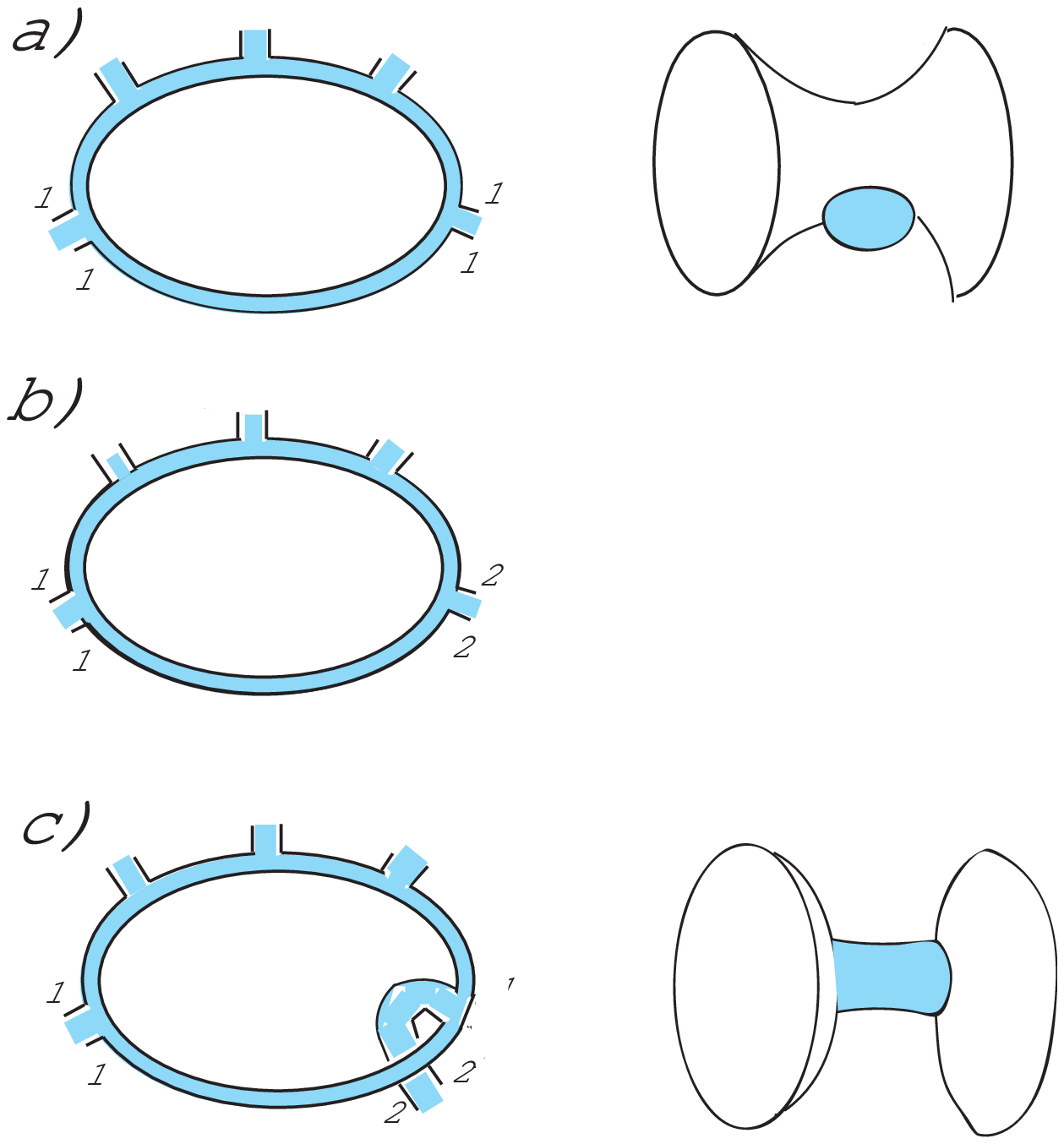}}

When $u=0$ becomes the minimum, there is even more we can say -
the radiative corrections, to all orders in perturbation theory,
are zero in the large $N$ limit.  In other words, there are no
planar diagrams in the expansion of the effective action about the
orthogonal $U(1)$'s configuration.  We argue as follows: take any
1PI loop diagram in the expansion of the effective action around
the classical orthogonal $U(1)$'s configuration and write it in
fatgraph notation (see \figOrtPert). Because the background is the
sum of two contributions coming from the two 't Hooft loops, we
can separate the external insertions into two types, I and II,
depending on which 't Hooft loop they came from. We can assume
that the first $U(1)$ is diag $(1, 0, \cdots, 0)$ and the second
is diag $(0,1,\cdots,0)$, so that the insertions in fatgraph
notation will be $(1,1)$ and $(2,2)$.  The key point now is that
any graph that has both type I and type II insertions cannot
possibly be planar - if we interpret the graphs as worldsheets
then type I and type II insertions cannot be put on the same
boundary (because there would have to be some propagator that
connects a type I and type II insertion, which is impossible, see
\figOrtPert (b)).  Note that this is true only for $s>s_0
(\lambda)$; for smaller $s$ we do not have orthogonal $U(1)$'s and
so it is possible to attach type I and type II insertions on the
same boundary.  This is suggestive because in the strong coupling
picture the corrections to the 't Hooft loop correlator are given
by F-string worldsheets ending on the D-brane; after the first
order transition these worldsheets (at leading order in $N$) turn
from disks to cylinders connecting the two D-string worldsheets
(see \figOrtPert (c)). Thus we see that $H(s)$ is constant, to all
orders in $\lambda$, for $s>s_0(\lambda)$. Another way to see this
is to use the fact that the state with orthogonal $U(1)$'s (which
are orthogonal only to leading order in $N$) is BPS to leading
order in $N$ and hence receives no radiative corrections.

We note that we could have attempted to do our entire analysis
with static temporal 't Hooft loops at finite temperature.  Here
we would be working with Euclidean $R^3 \times S^1$; \GrossGK\
have shown that there exists strong coupling first and second
order transitions in this case also.  The computational advantages
of working with this slightly simpler, static setup are offset by
the fact that the gauge theory on $R^3 \times S^1$ has bad
infrared divergences.  These occur at scales much longer than
those relevant for our problem, so presumably our analysis would
go through in some form.

\newsec {Discussion}

\ifig\figPhaseDiagram{Conjectured phase diagram in the $s$ -
$\lambda$ plane. The light solid line is the line of first order
transitions.  The dotted line is the line of second order
transitions in the metastable regime, the ``spinodal line".   The
heavy solid line is the line of second order transitions in the
stable regime emerging from the tricritical point.}
{\epsfxsize=9cm \epsfysize=7cm \epsfbox{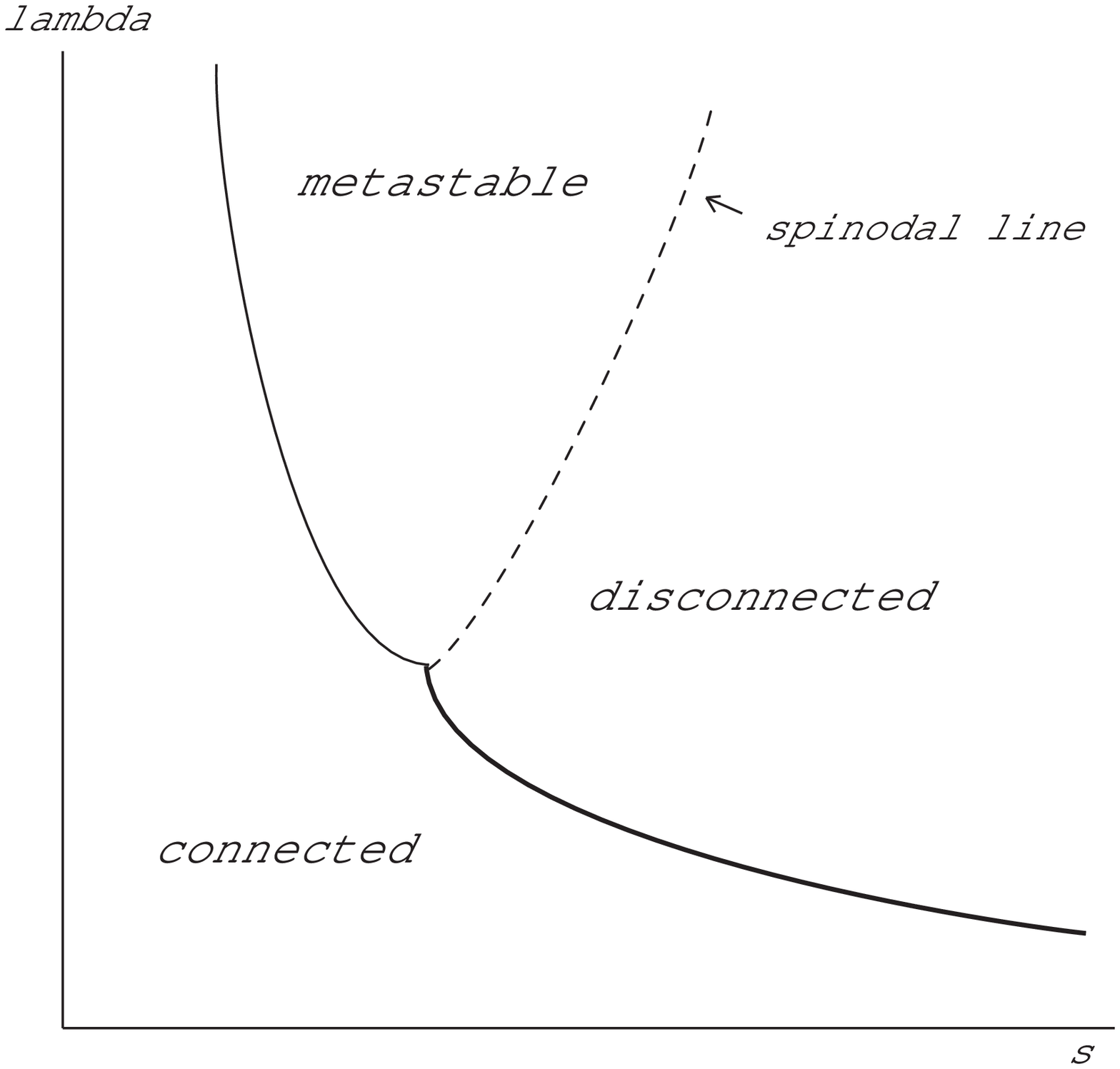}}

We have found that the singular behavior of $H(s)$ (the $\log$ of
the 't Hooft loop correlator) retains some information about the
phase transitions at strong coupling.  The second order transition
persists but the first order transition is absent.  Assuming a low
order polynomial approximation for $f$ as in mean field theory we
arrive at a natural conjecture for the phase diagram in the ($s$,
$\lambda$) plane (\figPhaseDiagram). We conjecture that the first
and second order transition lines  at strong coupling merge at a
tricritical point into a single line of second order transitions
located at $s_0 (\lambda)$.  Here the loci of both first and order
transitions vary analytically.   In general this need not be the
case because the location of a first order transition is
determined by a {\it real} condition:  that the real part of the
effective actions of two saddle points agree.

As discussed in the Introduction an important application of these
ideas is to the signature of the black hole singularity
\FidkowskiNF .  Here the branch cut at $t=0$ (Fig. 2) should
continue to large finite $\lambda$ as the scale of variation of
the coalescing geodesics is AdS scale, much larger than string
scale.  So this singularity should move analytically all the way
to small $\lambda$.

The operators that create wrapped D-branes are of the form ${\cal
O} = {\rm det}~ \phi^i$ where $\phi^i$ is one of the SYM scalar
fields \refs{\WittenXY, \BalasubramanianNH}. The combinatorial
formalism for evaluating correlators of ${\cal O}$ in perturbation
theory has been developed by Aharony, Antebi, Berkooz and Fishman
\AharonyND .  In the simpler case where the gauge group is
$SO(2N)$ and the operator ${\cal O} = {\rm Pfaff} ~\phi^i$ their
result is roughly the following (for large $N$):
\eqn\holey{\eqalign{\langle {\cal O} {\cal O} \rangle & \sim \int
{d \beta \over \beta} e^{2N f(\beta)} \cr f(\beta) &= \sum_{k \ge
1} \beta^k D_k - \half \log \beta }} Here$f(\beta)$ is a
generating function and $D_k$ denotes the sum over $k$ particle
irreducible diagrams. Parametrically $D_k \sim \lambda^{k-1}~.$ At
large $N$ \holey\ can be evaluated by finding saddle points of
$f(\beta)$. Roughly speaking one balances the ``entropy" of the
many different ways one can break a diagram up into different
subdiagrams against the ``energy" of choosing the largest $D_k$
values.   In a thermal system in Minkowski time the $D_k$ will
oscillate and decay \ArnoldGH\ allowing the possibility of
multiple saddle points exchanging dominance. These are good
candidates for the weak coupling image of the $t=0$ branch cut.

The dominant physics in such thermal correlators is efficiently
summarized by a Boltzmann transport equation dominated by two body
collisions occurring at a scattering rate $\lambda^2 T$ where $T$
is the temperature \ArnoldGH\ . It is plausible that the effects
of $k$ body collisions described by $D_k$ are suppressed relative
to the two body collisions by a factor $\lambda^{k-2}$, uniformly
in $t$. If this is the case  then $f(\beta)$ is convergent for all
$t$ and it is difficult to see an origin for the $t_c$
singularity, where $f$ should blow up. The $D_k$ should each be
finite at noncoincident points,  disfavoring the $\beta  \sim 0$
singularity which dominates for coincident points.   If these
tentative observations are correct then the $t_c$ singularity is
not present at small $\lambda$. By analyticity this would mean
that it has been smoothed out for all $\lambda$  less than
infinity. In other words, this signature of the black hole
singularity would be resolved purely by $\alpha'$ effects, even at
small $\alpha'$. This is surprising. We are continuing to
investigate these issues.

\newsec{Acknowledgements}
We would like to thank Ofer Aharony, Matthew Kleban, Hong Liu,
Xiao Liu, Juan Maldacena, Shiraz Minwalla, Hirosi Ooguri, Sergey
Prokushkin, Soo-Jong Rey,  Mohammed Sheikh-Jabbari, and Larry
Yaffe for helpful discussions. This work was supported in part by
NSF grant PHY-9870115 and the Stanford Institute for Theoretical
Physics. The work of L. F. was supported in part by a National
Science Foundation Graduate Research Fellowship.

\listrefs

\end